\documentclass[12pt]{article}
\usepackage[pdftex]{color, graphicx}
\usepackage{amsmath, amsfonts, amssymb, mathrsfs}
\usepackage{lscape}
\usepackage{float}
\usepackage{dcolumn}
\usepackage{hyperref, natbib}
\usepackage{amsthm}
\usepackage{bbm}
\usepackage{amsmath}
\usepackage{setspace}
\usepackage{subcaption}
\usepackage{float}

\date{}

\title{Assessing Ecosystem State Space Models: Identifiability and Estimation\vspace{-2ex}}
\author{Smith, J. W., Johnson, L. R., Thomas, R. Q.}
\author{Smith Jr., J. W.$^1$ \and Johnson, L. R.$^1$ \and Thomas, R. Q.$^2$}
\date{%
    \small
    $^1$Department of Statistics, Virginia Tech\\%
    $^2$Department of Forest Resources and Environmental Conservation, Virginia Tech\\[2ex]%
}

\begin{document}
\maketitle
\section*{Abstract}
Bayesian methods are increasingly being applied to parameterize mechanistic process models used in environmental prediction and forecasting. In particular, models describing ecosystem dynamics with multiple states that are linear and autoregressive at each step in time can be treated as statistical state space models. In this paper we examine this subset of ecosystem models, giving closed form Gibbs sampling updates for latent states and process precision parameters when process and observation errors are normally distributed. We use simulated data from an example model (DALECev) to assess the performance of parameter estimation and identifiability under scenarios of gaps in observations.  
We show that process precision estimates become unreliable as temporal gaps between observed state data increase. To improve estimates, particularly precisions, we introduce a method of tuning the timestep of the latent states to leverage higher-frequency driver information. 
Further, we show that data cloning is a suitable method for assessing parameter identifiability in this class of models. 
Overall, our study helps inform the application of state space models to ecological forecasting applications where 1) data are not available for all states and transfers at the operational timestep for the ecosystem model and 2) process uncertainty estimation is desired. \\
\textbf{Keywords:} Bayesian analysis, data cloning, ecological forecasting, MCMC 

\doublespacing
\vspace{-0.5cm}
\section{Introduction}

\vspace{-0.5cm}
Many ecological prediction and forecasting applications use mechanistic process-based models to simulate the dynamics of ecosystems \citep{Luo2011EcologicalFA}. These process models are typically discretizations of ordinary or partial differential equations describing how the system dynamics evolve over time and space. Further they are often linearly conditioned on the value at the previous time step. These models are especially important in applications where available data limits the use of empirical models or predictions are being made into novel conditions not captured in existing data. However, process-based models can be challenging to calibrate due to the large numbers of parameters, and thus robust uncertainty estimation is also difficult \citep{LUO2009}. 

While the importance of quantifying uncertainty in process models is recognized \citep{DIETZE2018}, it can be challenging to fully account for all important sources of uncertainty.  Research to date has largely focused on estimating and reducing uncertainty that arises from initial conditions, parameters, and observational noise, usually through data integration techniques \citep[see][for examples]{JIANG2018,WHITE2019,METEOLAKES}. Estimating and propagating these sources of uncertainty without process uncertainty assumes that the process model perfectly describes the temporal evolution of the ecosystem up to error in the collection of observations. To estimate process uncertainty, state space modeling frameworks \citep{Hamilton} are increasingly used to account for stochastic elements in the system evolution. Since many ecosystem process models already account for initial condition uncertainty, parameter uncertainty, and observational uncertainty, it is straightforward to convert them into a state space framework by adding an uncertainty structure to the underlying process model.

 The Bayesian state space paradigm is a well-suited approach to estimate distributions of parameters and latent states (model states that are not directly observed) in ecosystem models using observations. As a result it has seen a growing use in ecological forecasting applications \citep[see][for examples]{THOMAS2017,DOWD200339}. State space models treat all forecast terms as probability distributions and allow for more effective quantification, partitioning, and propagation of uncertainty in models. Prior distributions on parameters allow ecosystem scientists to enforce strict upper and lower bounds on parameters and incorporate biological information into the modeling process in a principled way. This focus on uncertainty and process precision estimation prompted us to choose a Bayesian framework over a point estimate focused method such as the Kalman filter \citep{Kalman}.   The parameter and latent state posterior distributions in Bayesian state space models are typically estimated using Markov Chain Monte Carlo (MCMC).   
 
The added flexibility of the state space model does come with drawbacks. Analyzing ecosystem models as state space models increases the number of parameters that require estimation (i.e., parameters describing the distribution of the process uncertainty) and adds the requirement to estimate latent states for each ecosystem model state. The addition of latent state estimation can add anywhere from tens to hundreds of thousands of additional parameters to estimate because ecosystem models commonly use a daily time step that requires an additional parameter per state for each day of the simulation. Additionally, a parameter is required for describing the process variance for each of the states in the model. Finally, data for ecosystem models may only be available at timescales that are less frequent than the model timestep (i.e., annual or greater). These large gaps between observed data may present a challenge to constraining both the latent states and process variances.   

 Furthermore, identifiability (or equifinality \citep{LUO2009}) is a common concern when using Bayesian methods to estimate parameters in ecosystem process models, which often have many highly correlated parameters. Although problems with parameter identifiability do not necessarily impair latent state or process variance estimation, these parameters correspond to properties of ecosystems and often have important physical interpretation. Therefore, it is crucial to ensure that they can be successfully identified and estimated.   
 
 Data cloning has been used  to assess identifiability of parameters in both phylogenetic models \citep{PONCIANO2012} and ecological models \citep{LELE2007}. DC is done by applying Bayesian inference to a dataset that is constructed by duplicating the initial dataset $r$ times. As $r$ increases, the resulting posterior parameter estimates approach the maximum likelihood solution. Data cloning can used to determine whether parameters are non-estimable or unidentifiable through a visual investigation of posterior plots with increasing values of $r$ \citep{PONCIANO2012}. Parameters are said to be non-estimable when there exist different parameter values $\theta_1, \theta_2, ..., \theta_n$ such that $\mathcal{L}(\theta_1 | X) = \mathcal{L}(\theta_2 | X) = ... = \mathcal{L}(\theta_n | X) = \text{argmax}_{\Theta} \mathcal{L}(\theta | X)$, i.e. there are multiple sets of parameter values that maximize the likelihood function \citep{LELE2010,RANNALA2002,ROTH1971,PONCIANO2012, Cole2020}. \cite{PONCIANO2012} introduce terminology for various situations where we have parameters in the model that are not estimable: non-separability, lack of information, nonidentifiable, and identifiable but not estimable. We use \citet{PONCIANO2012}'s definitions for these terms throughout the remainder of this paper and so we introduce these definitions here. Non-separability occurs when the model is structured such that it is not possible to separate parameters from one another, and may be due to parameter redundancy \citep[see also][]{Cole2020}. Parameters that are non-estimable due to non-separability are referred to as nonidentifiable (NI). Lack of information occurs when the dataset does not contain sufficient information about the parameters to estimate them, resulting in wide posterior distributions that have not been properly constrained by observed data. Parameters that are nonestimable due to lack of information are referred to as INE (identifiable but not estimable).  A combination of data cloning and effective sample size (ESS) estimation leads to a thorough analysis procedure for ecosystem state space models that has not commonly been applied.  

 To address challenges estimating latent states and parameters when applying Bayesian state space modeling frameworks to ecosystem models, we present a simulation study using a forest ecosystem state space model that predicts carbon cycling among multiple states (a.k.a. "stocks" in the carbon cycling model).  Using synthetic data with introduced data gaps, we address four questions focused on estimating variance parameters, latent states, and process parameters. 1) Under what temporal gaps in data does it become difficult to estimate the process variance? 2) Is it possible to recover the true latent states under different temporal gaps in data and under scenarios for availability of data describing the transfers among states (i.e., fluxes)? 3) Can we increase the sampling efficiency for the variance of latent states in these cases by disconnecting the time step of the latent states from the time step of the model? and 4) Can we determine when model parameters are identifiable using data cloning? Our study is designed to help inform the application of ecosystem state space models to ecological forecasting applications where process uncertainty estimation is desired, and where data is not available for all stocks and transfers at the ecosystem model operational time-step (i.e., a daily time-step), such as data collected through the U.S. National Ecological Observatory Network (NEON).

\vspace{-0.75cm}


\section{Methods}
\vspace{-0.5cm}
\subsection{Process Model}
We used the Data Assimilation Linked Ecosystem Carbon model designed for simulating forests composed of evergreen trees  \citep[DALECev,][]{WILLIAMSDALEC}.  It is a simple model describing carbon dynamics (Figure 1) and is similar to other ecosystem models used in data assimilation and carbon stock forecasting applications, for example  PnET \citep{PNET}, 3PG \citep{LANDSBERG1997}, TECOS \citep{TECOS}.  The model can be written as a set of equations that are approximately linear and autoregressive in time.  While DALECev has been widely used \citep{WILLIAMSDALEC, SMALLMAN, REFLEX, BLOOM} it is not traditionally fit as a state space model as we do here by adding a process error term into the model.  

\begin{figure}[H]
  \centering
  \includegraphics[scale = .4]{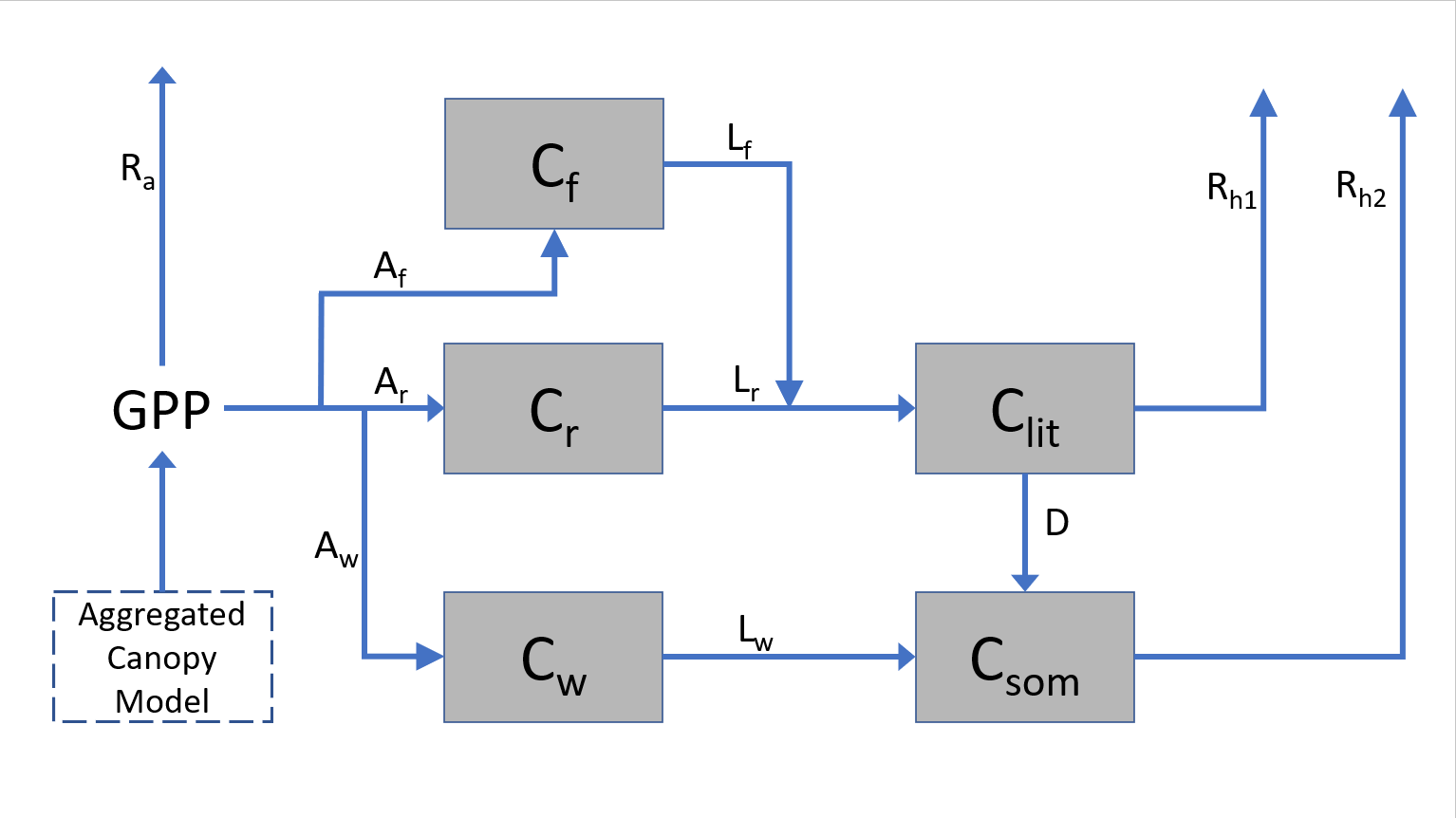}
  \label{fig:sfig1}
\caption{A schematic of the DALECev model with boxes denoting stocks of carbons and arrows denoting fluxes of carbon.}
\end{figure}

DALECev models the amount of carbon stored in five components within an evergreen forest ecosystem and is evaluated numerically at a daily time step, $t$. These five components, called stocks, include: carbon stored in foliage ($C_f ^{(t)}$); carbon stored in woody stems and coarse roots ($C_w^{(t)}$); carbon stored in fine roots ($C_r^{(t)}$); carbon stored in litter ($C_{lit}^{(t)}$); and carbon stored in soil organic matter ($C_{som}^{(t)}$). The DALECev model includes 11 process parameters, $p_i$, for $i \in 1 \dots 11$, where each parameter represents the daily rate of an ecological process (such as turnover, decomposition, or soil organic matter mineralization), an allocation of a particular flux (transfer rates between stocks), or a parameter used in the calculation of a flux. 

\begin{table}
\small
\begin{center}
 \begin{tabular}{|l|l |l|l|l|} 
 \hline
 Param. & Description & Lower & Upper &  Units \\ 
 \hline\hline
 $p_1$  &  Scaled daily decomposition rate & 1.1e-05 & 0.11 & day$^{-1}$ \\ 
 \hline
 $p_2$ &  Fraction of GPP respired & 0.2 & 0.7 & unitless \\
 \hline
 $p_3$ & Fraction of NPP allocated to foliage & 0.01 & 0.5 & unitless  \\
 \hline
 $p_4$  & Fraction of NPP after foliage allocation allocated to roots & 0.01 & 0.5 & unitless \\
 \hline
 $p_5$  & Daily turnover rate of foliage & 1e-04 & 0.1 & day$^{-1}$ \\
 \hline
 $p_6$  & Daily turnover rate of wood & 1e-06 & 0.01 & day$^{-1}$ \\
 \hline
 $p_7$ & Daily turnover rate of roots & 1e-06 & 0.01 & day$^{-1}$ \\
 \hline
 $p_8$  & Scaled daily mineralisation rate of litter & 1e-06 & 1 & day$^{-1}$ \\
 \hline
 $p_9$ & Daily mineralisation rate of soil organic matter & 1e-06 & 0.01  & day$^{-1}$\\
 \hline
 $p_{10}$ & Parameter for temperature dependent rate parameter & 0.05 & 0.2 & $^\circ$C $^{-1}$ \\
 \hline
 $p_{11}$  & Nitrogen use efficiency parameter & 2 & 20 & (gN gC$^{-1}$ d$^{-1})^{-1}$ \\
 \hline
\end{tabular}
\end{center}
\caption{Information on the physical interpretations, upper and lower bounds, and units for each of the eleven process parameters in DALECev. Parameter descriptions are taken from the REFLEX project supplemental material \citep{REFLEX}}
\end{table}

Fluxes represent a number of physical processes that move carbon through the ecosystem, including respiration ($R$), photosynthetic allocation ($A$), turnover ($L$), and transfer to another stock ($D$). The model uses a submodel, the Aggregated Canopy Model (ACM) from \cite{WILLIAMSACM}, to simulate the input of carbon through gross photosynthetic production (GPP; $G$ in Equations \ref{eq:Cft} -- \ref{eq:Crt}).  Following \citet{REFLEX}, all parameters in the ACM submodel were fixed except for $p_{11}$, therefore $G$ is a function of $p_{11}$ and meteorological driver inputs $\mathbf{D^{(t)}}$. $\mathbf{D^{(t)}}$ includes maximum and minimum temperatures, radiation, and atmospheric carbon dioxide for day $t$. For the DALECev model, a given carbon stock $C_s$ at time $t$ can be generically expressed as the carbon at time $t-1$ minus the respiration (carbon lost from the system) or transfer to another stock, plus carbon gained by through the allocation of photosynthesis (carbon gained from outside the system) or transfer from another stock, where for stock $s$:
\begin{align*}
C_s ^{(t)} = C_s ^{(t-1)} - L_s^{(t-1)} - R_s^{(t-1)} + A_s ^{(t-1)} \pm D_s ^{(t-1)} + \epsilon_{t-1,s}, \text{ } \epsilon_{t,s} \sim N(0, \phi_s).
\end{align*}
Here $\epsilon_{t-1,s}$ is an error term that acknowledges that our knowledge of the physical process is not perfect, and that there are normally-distributed stochastic deviations from the mean behavior.

Thus, our specific system of equations that simulates how model states change through time can be written as:
\begin{align}
\mathbb{E}[C_f ^{(t)}] &= C_f ^{(t-1)} - L_f ^{(t-1)}  + A_f ^{(t-1)} \nonumber  \\
& =  C_f^{(t-1)} - p_5 C_f^{(t-1)} + G(\mathbf{D^{(t)}}, p_{11})(1-p_2)p_3 
\label{eq:Cft} \\
\mathbb{E}[C_w ^{(t)}] &= C_w ^{(t-1)} - L_w ^{(t-1)} + A_w ^{(t-1)} \nonumber  \\
&=  C_w^{(t-1)} - p_6 C_w^{(t-1)} + G(\mathbf{D^{(t)}}, p_{11})(1-p_2)(1-p_3)(1-p_4) \label{eq:Cwt} 
\\
\mathbb{E}[C_r ^{(t)}] &= C_r ^{(t-1)} - L_r ^{(t-1)} + A_r ^{(t-1)} \nonumber  \\
&= C_r^{(t-1)} - p_7 C_r^{(t-1)} + G(\mathbf{D^{(t)}}, p_{11})(1-p_2)(1-p_3)p_4 \label{eq:Crt} 
\\
\mathbb{E}[C_{lit} ^{(t)}] &= C_{lit} ^{(t-1)} - R_{lit} ^{(t-1)} - D_{lit}^{(t-1)} + A_{lit} ^{(t-1)} \nonumber \\
&= C_{lit}^{(t-1)} - \Big [\frac{p_1 p_8 \exp(p_{10} \bar{T}^{(t)})}{2}  - \frac{p_1(1-p_8) \exp(p_{10} \bar{T}^{(t)})}{2} \Big] C_{lit}^{(t-1)} + \Big [p_5 C_f^{(t-1)} + p_7 C_r^{(t-1)} \Big] \label{eq:Clitt}
\\
\mathbb{E}[C_{som} ^{(t)}] &= C_{som} ^{(t-1)} - R_{som} ^{(t-1)} + D_{som}^{(t-1)} + A_{som} ^{(t-1)} \nonumber  \\
&= C_{som}^{(t-1)} - \frac{p_9}{2} \exp(p_{10} \bar{T}^{(t)} C_{som}^{(t-1)}+ \frac{(p_1 - p_1 p_8)}{2} \exp (p_{10} \bar{T}^{(t)})  C_{lit}^{(t-1)}  + p_6 C_w^{(t-1)} \label{eq:Csomt},
\end{align}
where $\bar{T}^{(t)}$ is the average temperature for day $t$. These updates are referred to as the process model, $f(\cdot)$. For any carbon stock $C$ the process model can be written in the form 
\begin{align}
\mathbb{E}[f(C^{(t-1)})] = A_t C^{(t-1)} + b_t \label{eq:TPALM},
\end{align}
where $A_t, b_t$ are coefficients that can vary with time. Any stocks that cannot be written in this way can be approximately written in this form using a linearization. 

The eleven different fluxes, the building blocks for Equations \ref{eq:Cft} -- \ref{eq:Csomt} in the DALECev model are:
\begin{align}
&R_a^{(t)} = G(\mathbf{D^{(t)}}, p_{11})p_2 \label{eq:fRa}\\
&A_f^{(t)} = G(\mathbf{D^{(t)}}, p_{11})(1-p_2)p_3 \label{eq:fAf}\\
&A_r^{(t)} =  G(\mathbf{D^{(t)}}, p_{11})(1-p_2)(1-p_3)p_4 \label{eq:fAr}\\
&A_w^{(t)} = G(\mathbf{D^{(t)}}, p_{11})(1-p_2)(1-p_3)(1-p_4) \label{eq:fAw}\\
&L_f^{(t)} = p_5 C_f ^{(t)} \label{eq:Lf}\\
&L_w^{(t)} = p_6 C_w ^{(t)} \label{eq:fLw}\\
&L_r^{(t)} = p_7 C_r ^{(t)} \label{eq:fLr}\\
&R_{lit} ^{(t)} = .5 \exp(p_{10} \bar{T}^{(t)}) p_1 p_8 C_{lit}^{(t)} \label{eq:fRh1}\\
&R_{som} ^{(t)} = .5 \exp(p_{10} \bar{T}^{(t)}) p_9 C_{som}^{(t)} \label{eq:fRh2}\\
&D_{lit}^{(t)} = .5 \exp(p_{10} \bar{T}^{(t)}) p_1(1-p_8) C_{lit}^{(t)} \label{eq:fDe}
\end{align}
These fluxes can be combined to form net fluxes. Unlike some of the individual fluxes or stocks, some of these net fluxes are measurable.  One important net flux is net ecosystem exchange ($\mathbf{NEE}$) that is measured using eddy covariance techniques \citep{Baldocchi} and deployed in many ecological observation networks \citep{Metzger}. $\mathbf{NEE}$ is the net of $G$ and Equations \ref{eq:fRa}, \ref{eq:fRh1}, and \ref{eq:fRh2} and is given by: 
\begin{align}
NEE^{(t)} = .5 \exp(p_{10} \bar{T}^{(t)}) p_1 p_8 C_{lit}^{(t)} + .5 \exp(p_{10} \bar{T}^{(t)}) p_9 C_{som}^{(t)} - G(\mathbf{D^{(t)}}, p_{11})(1-p_2) 
\end{align}
Additionally, soil respiration, $\mathbf{S_r}$, is a net flux that is commonly measured in ecosystem studies. $\mathbf{S_r}$ is the net of autotrophic respiration by roots (a component of Equation \ref{eq:fRa}) and heterotrophic respiration by soil micro-organisms(Equations \ref{eq:fRh1} and \ref{eq:fRh2}):
\begin{align}
S_r^{(t)} = .5 \exp(p_{10} \bar{T}^{(t)}) p_1 p_8 C_{lit}^{(t)} + .5 \exp(p_{10} \bar{T}^{(t)}) p_9 C_{som}^{(t)} - cG(\mathbf{D^{(t)}}, p_{11})p_2,\text{ } c \in [0,1]
\end{align}
For this study we have fixed the value of $c$ to be 0.3. In practice $c$ must either be specified or given a very strong prior, as it can be challenging to constrain by other available data.

\vspace{-0.5cm}

\subsection{State Space Model}

We estimate the stocks for the DALECev model using a state space model \citep{Hamilton, Petris, DurbinKoopman, am}. In the state space framework, we treat the five carbon stocks as components of the state vector, and the additional flux data collected on respiration, photosynthetic allocation, turnover, and transfers as operations on the state vector.  Let $\mathbf{C}$ denote the vector of carbon stocks from the model and $\mathbf{C}_{obs}$ denote the observations of the stock, with observations at a subset of time points $I \subset \{ 1, ..., T \}$. Then, Equations \ref{eq:Cft} - \ref{eq:Csomt} can be written compactly using matrix notation as:
\begin{align}
\mathbb{E}[\mathbf{C}^{(t)}] = M_t \mathbf{C}^{(t-1)} + \mathbf{P}^{(t)},
\end{align}
where 
$$
\footnotesize
\setlength{\arraycolsep}{2.5pt}
\medmuskip = 1mu 
\mathbf{C}^{(t)} = \begin{bmatrix}
           C_f ^{(t)} \\
           C_r ^{(t)} \\
           C_w ^{(t)} \\
           C_{lit} ^{(t)} \\
           C_{som} ^{(t)}
         \end{bmatrix}, 
M_t = \begin{bmatrix}
    (1 - p_5) & 0 & 0 & 0 & 0 \\
    0 & (1 - p_7) & 0 & 0 & 0 \\
    0 & 0 & (1 - p_6) & 0 & 0 \\
    p_5 & p_7 & 0 & (1 - \frac{p_1}{2} Q^{(t)}) & 0 \\ 
    0 & 0 & 0 & \frac{p_1 (1- p_8)}{2}Q^{(t)} & (1 - \frac{p_9}{2} Q ^{(t)}) 
  \end{bmatrix}, 
\mathbf{P}^{(t)} = \begin{bmatrix}
           G(\mathbf{D^{(t)}}, p_{11}) \psi_1 \\
           G(\mathbf{D^{(t)}}, p_{11}) \psi_2 \\
           G(\mathbf{D^{(t)}}, p_{11}) \psi_3 \\
           0 \\
           0
         \end{bmatrix},
$$
\normalsize with $Q^{(t)} = \exp(p_{10} \bar{T}^{(t)})$, $\psi_1 = (1 - p_2)p_3, \psi_2 = (1 - p_2)(1 - p_3)p_4$, $\psi_3 =  (1 - p_2)(1 - p_3)(1 - p_4)$. 

To relate our observations for an arbitrary carbon stock $C_{s, obs} ^{(t)}$ to the latent carbon stock $C_s ^{(t)}$, we assume the following relationship:
\begin{align*}
C_{s, obs} ^{(t)} = C_{s} ^{(t)} +\epsilon_{obs, t}, \text{ }t \in I, \text{ }\epsilon_{obs, t} \sim \mathcal{N}(0, \tau_s)
\end{align*}
In an ecological context, we are assuming that our observed carbon stock is normally distributed and unbiased, with a center at the true (latent) carbon stock, and a fixed precision $\tau_s$. Similarly to how adding the error term for the DALECev model was an acknowledgement of imperfect process knowledge, adding an error term for the observations is an acknowledgement of measurement error in the data that we observe. 

The state space model has two key assumptions: the state process is first order Markov, and the observation model is independent conditional on the latent states. Using normally distributed error terms for the process model and the observations, we can write these assumptions using the matrix notation above as:
\begin{align*}
\mathbf{C}^{(t)} | \mathbf{C}^{(t-1)} \sim \text{MVN} (M_t \mathbf{C} ^{(t-1)} + \mathbf{P}^{(t)}, \Phi), t = 1, \dots, T \\
\mathbf{C}^{(t)} _{obs} | \mathbf{C}^{(t)} \sim \text{MVN} (\mathbf{C} ^{(t)}, \tau), t \in I,
\end{align*}
where 
\begin{align*}
\footnotesize
\setlength{\arraycolsep}{2.5pt}
\medmuskip = 1mu 
\Phi = \begin{bmatrix}
    1/\phi_{Cf} & 0 & 0 & 0 & 0 \\
    0 & 1 / \phi_{Cr} & 0 & 0 & 0 \\
    0 & 0 & 1 / \phi_{Cw} & 0 & 0 \\
    0 & 0 & 0 & 1 / \phi _{Clit} & 0 \\ 
    0 & 0 & 0 & 0 & 1 / \phi_{Csom} 
  \end{bmatrix},
  \tau = \begin{bmatrix}
    1/\tau_{Cf} & 0 & 0 & 0 & 0 \\
    0 & 1 / \tau_{Cr} & 0 & 0 & 0 \\
    0 & 0 & 1 / \tau_{Cw} & 0 & 0 \\
    0 & 0 & 0 & 1 / \tau _{Clit} & 0 \\ 
    0 & 0 & 0 & 0 & 1 / \tau_{Csom} 
  \end{bmatrix}
\end{align*}
with all $\tau$ parameters assumed to be known. This assumption is not uncommon in terrestrial carbon models, as the measurement error is generally well understood. Fixing the measurement error can also lead to better estimation of other precisions, process parameters, and states  \citep[see][for a thorough discussion on this]{augermethe2016}. The combination of a linear process model, normally distributed process error, and normally distributed measurement error means that we are fitting DALECev as a Normal Dynamic Linear Model (NDLM) \citep{HarrisonWest}.

The fluxes are modelled with a simple observation model. For a given flux $F_j$, with flux data collected at a subset $I_j \subset \{ 1, ..., T \}$ and observation $F_{j, obs}$, we assume the relationship 
\begin{align}
F_{j, obs} ^{(t)} | F_{j} ^{(t)} \sim \mathcal{N} (F_j ^{(t)}, \delta_j ), \text{ } t \in I_j,
\end{align}
where $\delta_j$ is a known precision. This specification for the fluxes assumes that flux observations are unbiased but contain measurement error. Each individual flux has different observation time points $I_j$ to account for the fact fluxes are measured using different methods, and the methods may work on different timescales, as well as giving flexibility in the case of data collection failure. 

With models assigned for our physical process, observations, and fluxes, we can write the full likelihood of the model as:
\begin{align}
\mathscr{L}(\mathbf{C}^{(1:T)}, \Phi, p_{1:11} | \mathbf{D}^{(1:T)}, \tau, \delta_{1:J}, \mathbf{C}_{obs}, \mathbf{F}_{obs}) =& \prod_{t = 1}^T \text{MVN} (M_t \mathbf{C} ^{(t)} + \mathbf{P}^{(t)}, \Phi) \nonumber\\ 
\times &\prod_{t \in I} \text{MVN} (\mathbf{C} ^{(t-1)}, \tau) \nonumber \\ 
\times &\Bigg( \prod_{j = 1}^J \prod_{t \in I_j}  \mathcal{N} (F_j ^{(t)}, \delta_j ) \Bigg)
\label{eq:likelihood}
\end{align}

Many prior studies using terrestrial carbon models do not fit process error (see \cite{JIANG2018} for example), instead using observational uncertainty in the estimation of the process model likelihood. The state space approach used here is designed specifically to help isolate the process uncertainty from observational and parameter uncertainty.  This partitioning of uncertainty is critical in understanding the system because no one source or type of uncertainty is likely to dominate total model uncertainty across all of ecological applications \citep{DIETZE2018} and they influence forecast uncertainty in different ways, e.g., process uncertainty propagates from one timestep to another while observation uncertainty does not. The Bayesian state space paradigm outlined here allows for quantification of multiple sources of uncertainty (process, initial condition, and observation) in the context of temporal gaps in observations, and the state space model gives a natural setting to leverage observed data with process based models. 

\subsection{Inference for parameters and latent states}
We estimate the stocks, process precisions, and process parameters for the DALECev model using a Bayesian state space model \citep{Hamilton, DurbinKoopman, Petris, am}. Process parameters, process precisions, and latent states were estimated using MCMC \citep{CasellaRobert}. MCMC is a flexible method that uses Markov chains to generate samples of the parameters from their posterior distribution. Parameter uncertainty is usually inherently included in MCMC methods, and the samples from the posterior can be used to generate credible intervals for where the true values of the parameters lie. 
Equipped with the model likelihood from \ref{eq:likelihood}, we need to choose prior distributions for the process parameters, process precisions, and initial conditions. Prior choices for the process parameters are assumed to be uniform with limits informed by the range of values gathered from expert opinion in the Reflex project supplemental material \citep{REFLEX} to reflect the change in sites to the Talladega National Forest (see description below). The values for $p^{(L)}$ and $p^{(U)}$ can be found in Table 1. Each process precision was given a univariate conjugate Jeffreys prior \citep{JEFFREYS1946}, to allow for closed form Gibbs sampling of the process precision parameters. Thus the priors are given by
\begin{align}
&p_i \sim \mathrm{Unif}(p_i ^{(L)}, p_i ^{(U)}) \\
&C_k ^{(0)} \sim \mathcal{N}(\mu_k ^{(0)}, \phi_k ^{(0)}) \\
&\pi(\phi_k) \propto \frac{1}{\phi_k}.
\end{align}

Building on the likelihood and priors, we can derive the full conditional distributions for all latent stocks and and precision parameters. The full condition distributions for latent carbon stocks at interior (between the initial and final) time steps with observed data can be written as:
\begin{align}
\pi(C_k ^{(t)} | C_{-k}^{(t)}, C_{k, obs}^{(t)}, \cdot) \sim \mathcal{N}\Big(\frac{\phi_k (A_t C_k ^{(t-1)} + b_t + A_{t+1} (C_k ^{(t+1)} - b_{t+1})) + \tau_k C_{k, obs} ^{(t)}}{\phi_k (1 + A_{t+1} ^2) + \tau_k}, \phi_k (1 + A_{t+1} ^2) + \tau_k \Big). \label{eq:Gibbs1}
\end{align}
The full condition distributions for latent carbon stocks at interior time steps without observed data at those time points can be written as:
\begin{align}
\pi(C_k ^{(t)} | C_{-k}^{(t)}, \cdot) \sim \mathcal{N}\Big(\frac{\phi_k (A_t C_k ^{(t-1)} + b_t + A_{t+1} (C_k ^{(t+1)} - b_{t+1}))}{\phi_k (1 + A_{t+1} ^2)}, \phi_k (1 + A_{t+1} ^2)\Big).
\end{align}
The full conditional distributions for the initial latent state and final latent states are:
\begin{align}
\pi(C_k ^{(0)}| \cdot) &\sim \mathcal{N} \left( \frac{\phi_k (A_1C_k ^{(1)} - A_1 b_1 + \phi_k ^{(0)} \mu _k ^{(0)})}{\phi_k A_1 ^2 + \phi_k ^{(0)}} , \phi_k A_1 ^2 + \phi_k ^{(0)}\right) \\
\pi(C_k ^{(T)}| \cdot) &\sim \mathcal{N} \left(\frac{\phi_k (A_T C_k ^{(T-1)} +b_t) + \mathbbm{1}_{T \in I} (\tau_k C_{k,obs} ^{(T)})}{\phi_k + \mathbbm{1}_{T \in I} (\tau_k)}, \phi_k + \mathbbm{1}_{T \in I} (\tau_k) \right) ,
\end{align}
where $\mathbbm{1}_{T \in I}$ is an indicator function that is equal to 1 if there is an observation for $C_k$ at the final time point, and 0 otherwise.  Finally, the full conditional distributions for the precisions are:
\begin{align}
\pi (\phi_k | \cdot) \sim \Gamma\left( \frac{T}{2}, \frac{1}{2} \sum_{t = 1}^T (C_k ^{(t)} - f(C_k ^{(t-1)}) )^2 \right), \label{eq:Gibbslast}
\end{align}
 where $\Gamma$ is the univariate gamma distribution using the rate parameterization. 
 
 We estimated the posterior distributions of the parameters using MCMC in the {\sf R} programming language \citep{R}, specifically the Random Walk Metropolis Hastings (RWMH) (\cite{METROPOLIS}; \cite{HASTINGS}) with a block updating algorithm. We sampled process parameters using a univariate proposal distribution during the initial burn-in phase (2,000 iterations). After burn in, we sampled parameters for 500 iterations where we jointly sampled highly correlated process parameters using a truncated normal proposal that accounts for their covariance. We recalculated the empirical covariances used in the block updates every 500 iterations. We updated the analytic full conditional distributions given in Equations \ref{eq:Gibbs1} - \ref{eq:Gibbslast} for the latent states, process precisions, and initial conditions using a Gibbs sampler \citep{GemanAndGeman}. Process parameters were updated using the RWMH. Including burn-in, 10,000 total posterior samples were collected.
 
Careful attention was paid to the initialization of the latent states for the MCMC. We initialized the unobserved latent states in one of two ways. For small data gaps a Gaussian process \citep{KENNEDYOHAGAN} was fit using the \texttt{laGP} package \citep{laGP} in {\sf R} \citep{R}. For large data gaps a particle filter was used (as implemented in \cite{GORDON}). The Gaussian process has the advantage of not requiring initial estimates for the process parameters and process precisions at the cost of not incorporating knowledge of the ecosystem model and certain properties of the carbon stocks, such as positivity.  However, with low observation data frequency, the particle filters use of the process model helps to provide better initial estimates of latent states, speeding up convergence. To initialize the particle filter, maximum likelihood estimates for the process parameters are computed from the flux data and stock observations. In the event of minimal flux data, a space filling design, in this case a Latin Hypercube sample (LHS) \citep{mckay_LHS}, of the process parameters is used to create sets of process parameters. These sets of process parameters are then run through a bootstrap particle filter using the initial condition distributions, with the process precision set to a fixed percentage of the mean of the initial condition distribution. The combination of process parameters and latent states with the highest likelihood from the corresponding LHS and bootstrap filtering are then used for their respective initializations.

\vspace{-0.5cm}

\subsection{Simulation study design}

We designed a simulation study to evaluate the ability of standard MCMC methods to estimate process precisions, latent states, and process parameters for the DALECev model, and to identify and address potential problems that may arise when using these methods for the types of (sometimes sparse) data that are available. More specifically, we had three primary objectives. The first was to look at when we can expect precision estimates to break down, and whether changing the timestep of the latent states can help to alleviate issues with estimation. The second was to examine how well the model recovers the latent carbon stocks with only annual observed data, which is the type of data that we can expect to collect to parameterize models like DALECev. Third we wanted to assess parameter identifiability (via data cloning) when fitting the models to different data that are available and use this information to help inform data collection schemes. 
 
We began by generating a set of synthetic datasets for use in our analysis.  Our simulation study was created to emulate conditions at the Talladega National Forest in Alabama, U.S.A (32.95046$^{\circ}$  N, -87.39327$^{\circ}$  W).  We chose this site for two reasons. First, the site has a canopy dominated by evergreen tree species (longleaf pine (\textit{Pinus palustris}), loblolly pine (\textit{Pinus taeda}), and slash pine (\textit{Pinus elliotti})) that matches the canopy type expected by the DALECev model. Second, the site is part of the National Ecological Observatory Network (NEON) and thus has ongoing data collection that can be used in future applications of the methods described here. For the synthetic data set, initial conditions and driver data for the carbon stocks were derived from NEON data \cite{NEON_data}, with specified initial mean and initial uncertainty. Process parameter values for simulations were chosen such that carbon stock data, leaf area index (LAI), and NEE were reflective of what would be expected at Talladega. Exact values for the process parameters used in the simulations can be found in Table \ref{tab:sim_values}. Random initial conditions were drawn from their respective prior distributions. At each time step process noise is added to the states, with observational noise added to the latent states at the end of the model run to create a dataset of observations. Data gaps for synthetic datasets were created by removing observations that are not in the analysis time step. 
 
\begin{table}
\small
\begin{center}
 \begin{tabular}{|l|l |l|l|} 
 \hline
 Param. & Value & Description & Units \\ [0.5ex] 
 \hline\hline
 $p_1$  & .002 & Scaled daily decomposition rate & day$^{-1}$\\ 
 \hline
 $p_2$ & .27 & Fraction of GPP respired & unitless \\
 \hline
 $p_3$ & .15 & Fraction of NPP allocated to foliage & unitless \\
 \hline
 $p_4$ & .33 & Fraction of NPP after foliage allocation allocated to roots & unitless \\
 \hline
 $p_5$ & .00137 & Daily turnover rate of foliage & day$^{-1}$ \\
 \hline
 $p_6$ & 1.1e-04 & Daily turnover rate of wood & day$^{-1}$ \\
 \hline
 $p_7$ & .00137 & Daily turnover rate of roots & day$^{-1}$ \\
 \hline
 $p_8$ & .1  & Scaled daily mineralization rate of litter & day$^{-1}$ \\
 \hline
 $p_9$ & 1.096e-05  & Daily mineralization rate of soil organic matter & day$^{-1}$\\
 \hline
 $p_{10}$ & .1725 & Parameter for temperature dependent rate parameter & $^\circ$C $^{-1}$ \\
 \hline
 $p_{11}$ & 3  & Nitrogen use efficiency parameter & (gN gC$^{-1}$ d$^{-1})^{-1}$ \\
 \hline
\end{tabular}
\end{center}
\caption{Parameter values used to generate synthetic data sets of carbon stocks and fluxes using the DALECev model. Parameter values were chosen such that carbon stock data, leaf area index (LAI), and NEE were reflective of what would be expected at Talladega. Parameter descriptions are taken from the REFLEX project supplemental material \citep{REFLEX}}
\label{tab:sim_values}
\end{table}




\vspace{-0.5cm}

\vspace{-0.5cm}

\subsection{Effects on estimates of precision and the latent states of varying the state process time resolution}
The ability to estimate process precision is crucial in ecosystem models, as that is often the main source of uncertainty \citep{DIETZE2018, THOMAS2017}. Poor estimation of process precision may lead to more uncertainty in estimates of process parameters and of latent states, which can then affect forecasts and make them unreliable. For models like DALECev, gaps between observations of the states are commonly greater than 1 year, a much slower time scale than the assumed process dynamics, resulting in many unobserved states. In order to reliably apply DALECev in practice, process precision and latent state estimation should be robust to annual or longer data gaps.  

  To determine when the estimation of process precision becomes difficult, we examined three different observation scenarios: daily state observations, monthly state observations, and yearly state observations, each with daily flux observations. Initial conditions were drawn from the prior distributions, and driver data from the site was used to run each model for two years. We repeated the generation of the synthetic dataset 15 times. For each dataset, observations were removed to introduce synthetic data gaps that matched the different observation scenarios. 
 
  We evaluate the performance of precision estimation using the effective sample size (ESS) \citep{robertcasellamontecarloinr} of the Markov chains for the precision parameters as a metric. Samples from Markov chains are auto-correlated, meaning that they are correlated with previous samples in the chain. The ESS gives us a metric for determining how efficiently our chain has estimated parameters/states, with low ESS informing us that our samples are highly correlated and to proceed cautiously with using the posterior samples for prediction or interpretation.  
 
 To evaluate ESS under different gaps in data, we used MCMC (as described above) to estimate posterior parameter distributions for each dataset and analyzed the process precision MCMC chains using ESS as a measure of sampling efficiency. We identified the data gaps where a large degradation in sampling efficiency occurred.

It also can be difficult to obtain information about latent states when there are large gaps between observed data points.  We explored whether changing the latent state timestep is a solution for alleviating problems with low effective sample sizes without introducing estimation problems for process parameters. These problems may arise from differences in the flux data and observation data likelihoods having different time steps. To analyze these differences, the model was run with a daily latent state resolution and was compared to the model being run with a monthly latent state resolution. 
 
Consider an NDLM on a daily process model for an arbitrary carbon stock $C$, with a process model of the form:
\begin{align}
C^{(t)} = A_t C^{(t-1)} + b_t. \label{eq:process}
\end{align}
Let $T^* = \{ t_i ,~~ i = 1, .., I\}$ be a proper subset of the timesteps of the model. For an NDLM with a process model of the form in Equation (\ref{eq:process}), state transitions can be rewritten in the form
\begin{align}
C^{(t_i)} | C^{(t_{i-1})} \sim \mathcal{N} \Bigg( \Big(\prod_{j = t_{i-1}+1} ^{t_i} A_j \Big) C ^{(t_{i-1}) } + \sum_{k = t_{i-1}+1 } ^ {t_{i}} \Big(\prod_{m = k}^{t_i} A_m \Big)b_m  + b_{t_i}, \frac{\phi_s}{1 + \sum_{k = t_{i-1}+1 } ^ {t_{i}} \Big(\prod_{m = k}^{t_i} A_m \Big)} \Bigg). \label{eq: state_trans}
\end{align}
This allows the stocks of the model to operate on a different time step than the fluxes, so that daily flux information can be used without requiring estimation of a large number of latent states that have very little data to constrain them. It also gives the model more flexibility, allowing models to change time steps for inference purposes, to decrease computational costs, and to allow for varying time steps across the stocks themselves. For values for $A$ that are constant or similar through time and approximately uniformly spaced entries of $T^*$, we may treat the precision in Equation (\ref{eq: state_trans}) as a fixed value $\phi^*$. Using this approach, we follow the advice given in \cite{am}, which says "make simplifying assumptions when data are limited". 

 To examine how sampling efficiency is influenced by changing the time step at which latent states are estimated, we used two different synthetic datasets. The first synthetic dataset is a set of of 30 synthetic time-series that have monthly carbon stock observations and daily flux observations. The second synthetic dataset is a set of 30 synthetic time-series that have annual carbon stock observations and daily flux observations. Each synthetic dataset was analyzed using both a daily time step and a monthly time step in the state process model. We compared the effective sample sizes of process precisions between the different observation regimes to assess whether changing the time step of latent state estimation increases sampling efficiency of process precision parameters.

\vspace{-0.5cm}

\subsection{Impact of observation gaps  on estimation of latent states}
While efficient estimates of process precision are important, the primary goal of an ecological state space model is to accurately track the evolution of the latent states through time. This involves both latent state estimation and estimation of the underlying process parameters. A particular focus was placed on annual carbon stock availability, as it is the most realistic case when working with actual data.  

 Latent state estimation was assessed by creating synthetic datasets with daily, monthly, and annual carbon stock data availability. Posterior estimates of the la. The latent state estimates were compared to their true (synthetic) values. Synthetic datasets were analyzed using both a daily model time step and the monthly model time step for the latent states. Coverage rates of latent state posterior distributions were assessed to confirm that they reach roughly the nominal rate.  This was achieved by checking if the true latent state values were contained in the 95\% highest posterior density intervals, and taking the average for each stock over all time steps.
 
  We also assessed convergence of process parameters. MCMC inference was conducted on identical datasets with four different starting conditions for the process parameters. We then used the Gelman Rubin test statistic \citep{gelman_rubin} across the MCMC chains to assess convergence of process parameters. 
  
  \vspace{-0.5cm}

\subsection{Parameter identifiability under differing data availability} 

 Identifiability of parameters was assessed using data cloning to analyze three synthetic datasets with annual carbon stock observations. Each dataset had different levels of flux data observations: 1) all fluxes observed; 2) fluxes available from NEON with GPP data; 3) only fluxes available from NEON with NEE data. These were chosen so that we could compare the ideal case to data that would be commonly available for terrestrial carbon models. Our MCMC inference proceedure was performed on each of the synthetic datasets with $r = 1, 5, 25$ data cloning replicates, and posterior distributions of $p_2, p_3, p_4, p_{11}$ were analyzed across datasets and data cloning replicates.  

 

Revisiting Equations (\ref{eq:fAr}) and (\ref{eq:fAw}) from Section I, we see that $A_r$ and $A_w$ give additional information for $p_2, p_3, \text{ and } p_4$, a set of parameters that are highly correlated due to their entanglement in the carbon update equations. The absence of one or more of these fluxes, like in the case of using NEON data only, may make these parameters difficult to identify. For scientists, a data cloning analysis can serve more purposes than just assessing identifiability of parameters. Simulated data can be used a priori to determine what data is most important to collect in their experiment.

\vspace{-0.5cm}

\section{Results}
\vspace{-0.5cm}

\subsection{Estimating process precision under data gaps}
\begin{figure}[H]
  \centering
  \includegraphics[scale = .6]{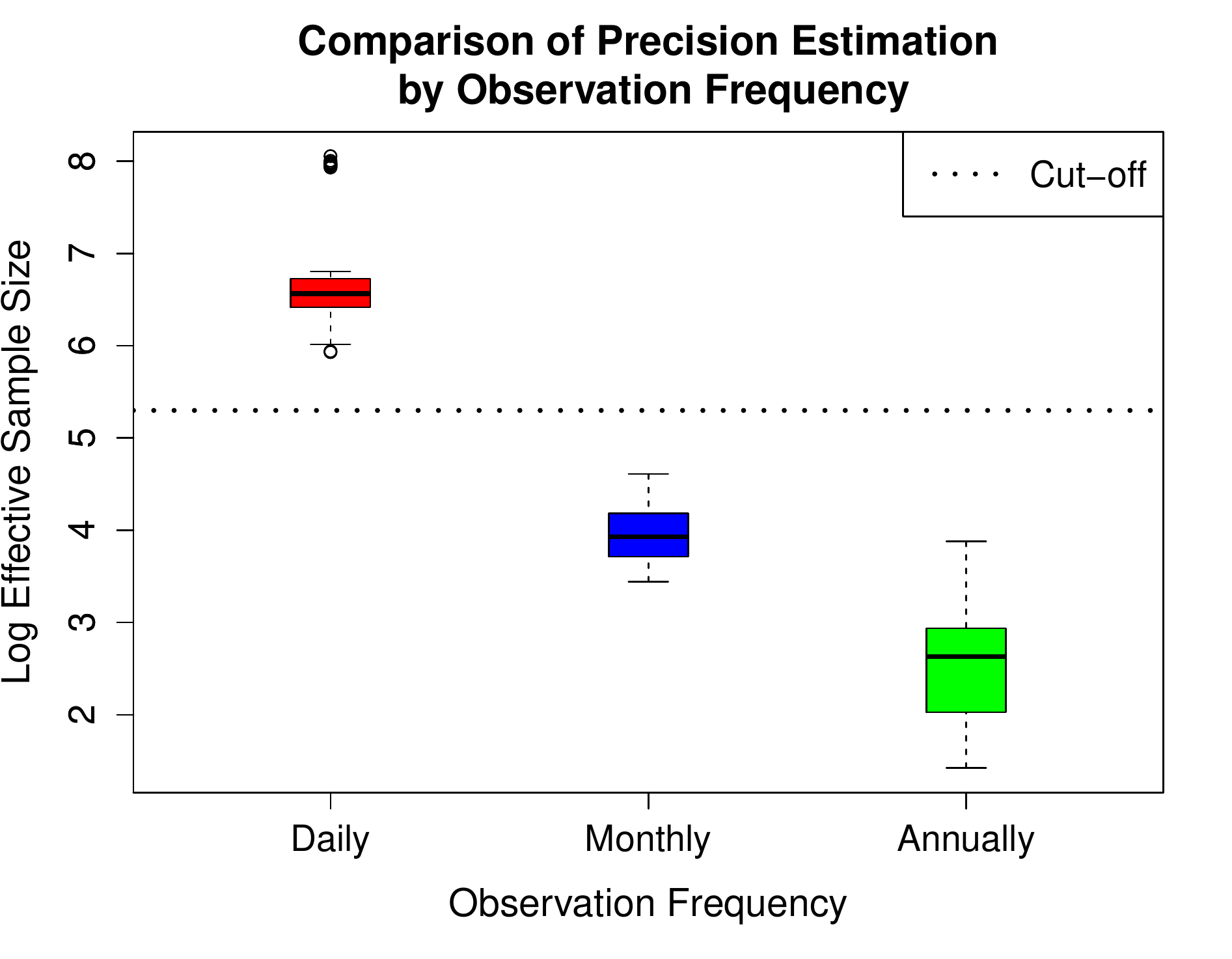}
\caption{A plot of log average ESS of estimated precision parameters versus observation frequency. In each case we used the daily time step model and averaged over fifteen MCMC runs of 10,000 iterations of each model configuration. The dotted line indicates the cut-off of an ESS of 200 (5.30). }
\label{fig:GAP_ESS}
\end{figure}

As would be expected, we found that increasing the time between observations leads to a noticeable decrease in MCMC sampling efficiency (Figure \ref{fig:GAP_ESS}). The criteria we used to assess whether or not the posterior samples size was large enough for for inference was an ESS of 200 or higher. By this metric, both monthly and annual observation data give ESS's that are too low to make credible posterior inference.  
In particular, yearly observed data produces precision estimates 
with very low effective sample sizes. This is particularly concerning, as annual observations are the most realistic of the three data availability scenarios we explored. This finding motivates the need to alter the sampling methods in order to increase the sampling efficiency of these precision parameters. 

\vspace{-0.5cm}

\subsection{Effects on estimates of precision and the latent states of varying the state process time resolution}
\begin{figure}[H]
  \centering
  \includegraphics[scale = .6]{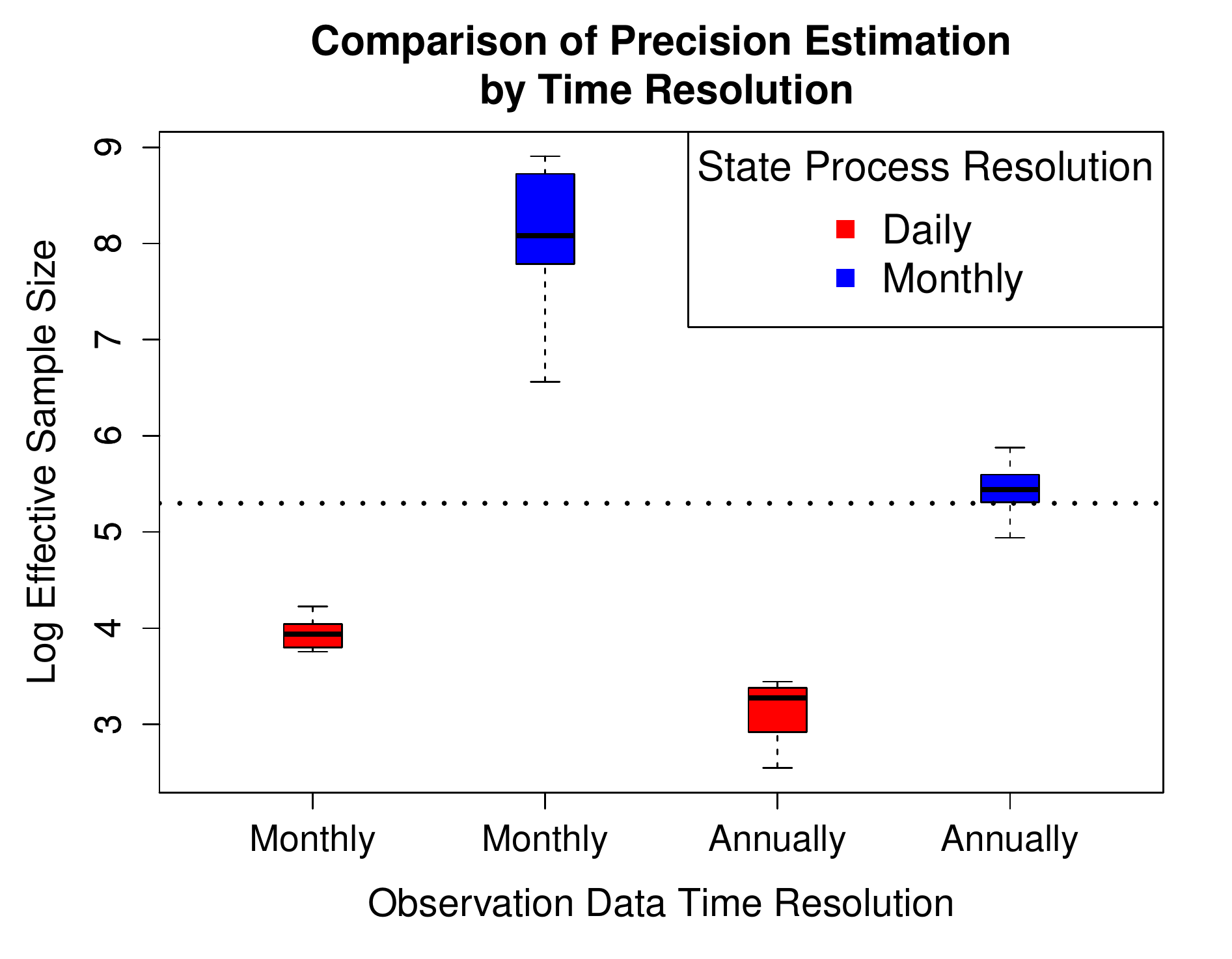}
\caption{A plot of observation frequency vs. log average effective sample sizes for the estimated precision parameters (inverse variance) for different observation collection regimes using the daily step and the monthly step models. In each case we averaged over fifteen MCMC runs of each model configuration. The dotted line indicates the cut-off of an ESS of 200 (5.30). Precision log ESS is pooled over all five precisions corresponding to their respective carbon stocks.}
\label{fig:ESS_plot2}
\end{figure}

In Figure \ref{fig:ESS_plot2} we compare the effective sample size of several simulations with different timesteps. We found that changing the model timestep from daily to monthly improves the estimation of process precision for both monthly and annual data gaps. This showcases the increased performance for the change of the latent state timestep as observation sparsity increases. For both monthly and annual data observations, the monthly timestep model performs significantly better. As expected, effective sample sizes for precisions decrease across timestep type as observation sparsity increases.  

\vspace{-0.5cm}

\begin{figure}[H]
	\centering
  \includegraphics[scale = .7]{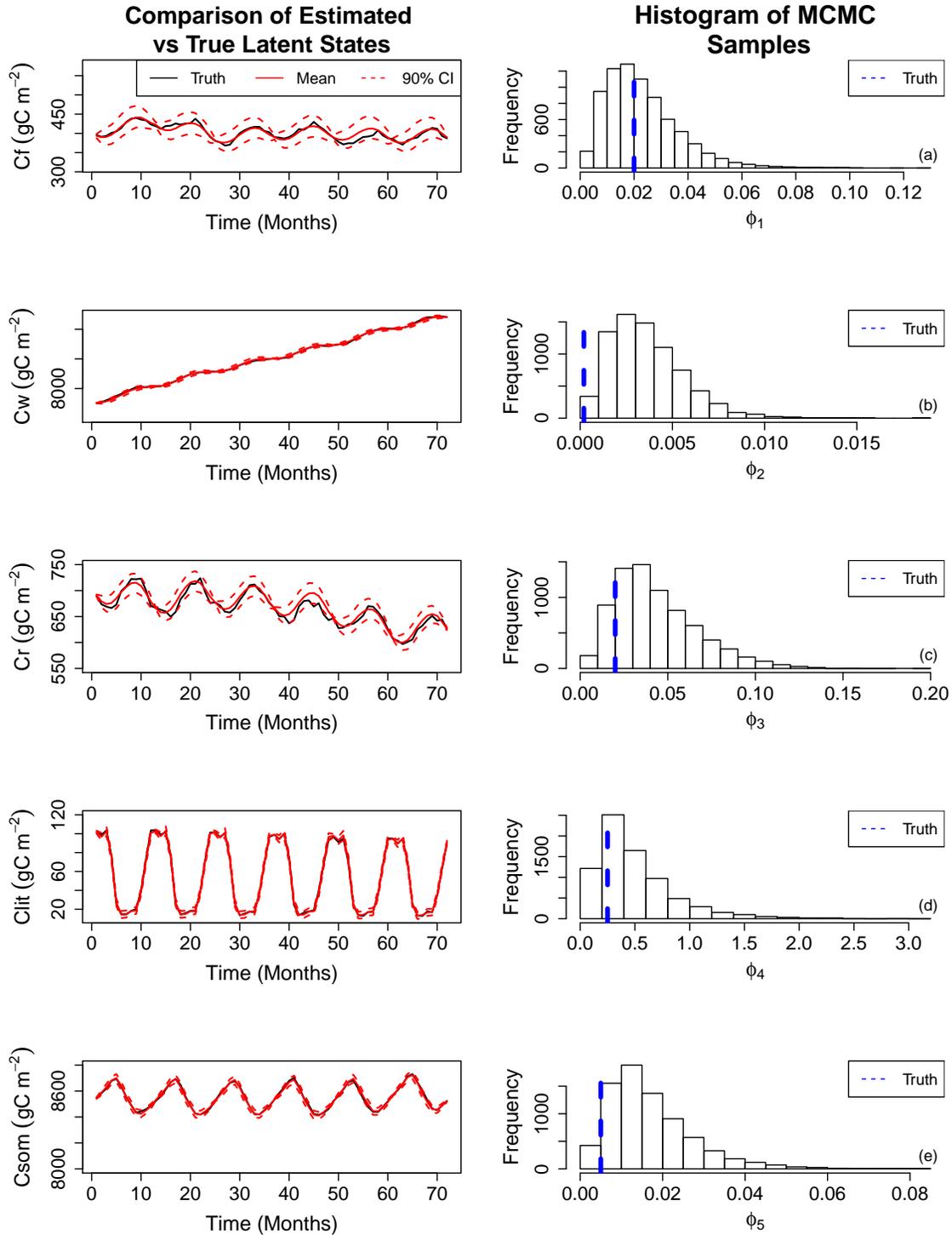}
\caption{(a-e) Estimates for latent states for a monthly timestep with annual data observations (Left) along with histograms of corresponding marginal estimated process precision post burn-in (Right). Each panel corresponds to a particular carbon stock: $C_f$ represents foliage carbon, $C_w$ represents wood carbon, $C_r$ represents root carbon, $C_{lit}$ represents litter carbon, and $C_{som}$ represents soil organic matter carbon.}  
\label{fig:latent_plot}
\end{figure}

An important goal of analyzing ecosystem process models from a state space framework is to track and predict the evolution of latent states through time. In Figure \ref{fig:latent_plot} we show posterior latent state estimates for carbon stock data observed annually, with all flux data observed daily and the model running on a monthly time step. 

 In Table \ref{tab:cvg} we show the average performance of the monthly time step model over 30 different model runs. Coverage rate of posterior credible intervals for the latent states are used to quantify how well latent states are being tracked. For each latent variable, the average probability that the true process precision lies within the 95\% credible interval is also given.  

\begin{table}
\begin{center}
 \begin{tabular}{| c | c | c |} 
 \hline
 Stock variable& Avg.~Coverage & Precision (\%) \\ 
 \hline
 $\mathbf{C_f}$  & .966 & 100 \\ 
 \hline
 $\mathbf{C_w}$ & .935 & 100  \\
 \hline
 $\mathbf{C_r}$ & .950 & 100 \\
 \hline
 $\mathbf{C_{lit}}$ & .990 & 92.9  \\
 \hline
 $\mathbf{C_{som}}$ & .987 & 100  \\ 
\hline
\end{tabular}
\end{center}
\caption{Average empirical coverage and percent of precisions contained in the 95\% HPD credible interval, averaged over 30 model runs }\label{tb:coverage}
\label{tab:cvg}
\end{table}

  Nearly all of the marginal posterior distributions of the process precisions in Figure \ref{fig:latent_plot} show right skewness. This may come from a trade-off between model precisions, where one precision being over-estimated leads to another precision being under-estimated. Another possibility is that our independence assumption between carbon stocks is being violated. Some carbon stock updates, including $C_{lit}$ and $C_{som}$, involve other carbon stocks. This may be causing an underestimation of precision, which in turn affects the estimation of other process precision parameters. 
  \textbf{}

\vspace{-0.5cm}

\subsection{Parameter identifiability under differing data availability}
Our analysis of the data cloning results involves the following set of considerations. First, for identifiable parameters, we expect that as $r$ increases, the variance of the resulting estimate decreases. This can be seen when the resulting posteriors grow tighter around the mean as values of $r$ get larger. Second, identifiable but non-estimable parameters (INE) are parameters that may be identifiable, but do not have a necessary amount of data to tease out the precise values. These are characterized by relatively flat posterior distributions \citep{PONCIANO2009}. Finally, parameters that are non identifiable (NI) tend to have multi-modal posterior distributions, indicating that there are several values of the parameter that produce a high value of the likelihood. Functions of multiple non identifiable parameters can be estimable, but the individual parameters themselves are not. See \citep{PONCIANO2012} for a simple example.

\begin{figure}[H]
\centering
\includegraphics[scale = .65]{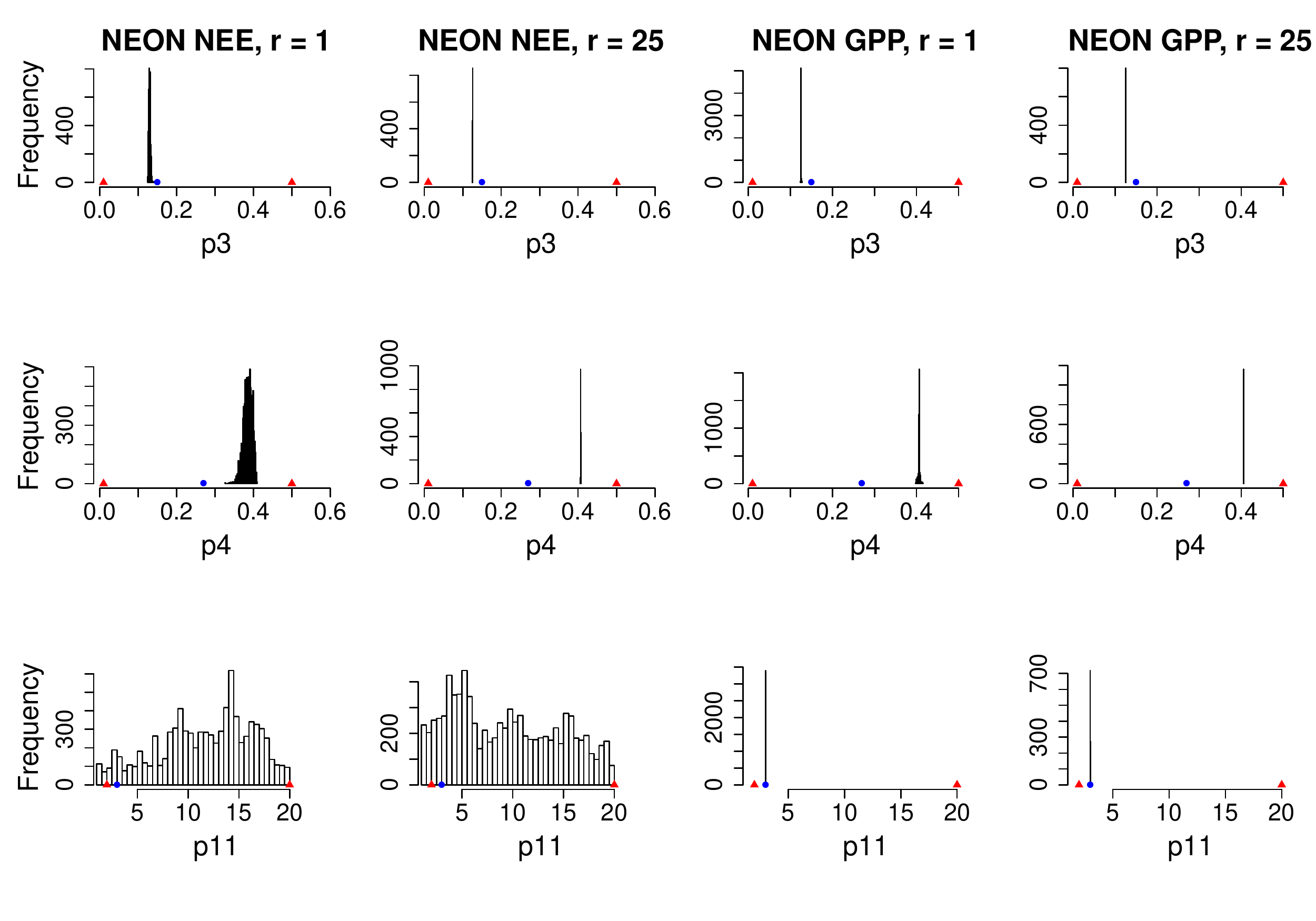}
\caption{Results of data cloning for selected process parameters under the two NEON data flux scenarios, with 1 data cloning replicate and 25 data cloning replicates. Marks on the x axis denote the upper and lower bounds of the uniform priors given to the process parameters.}
\label{fig:DC_plot}
\end{figure}

We found that data cloning served as an effective way to assess identifiability of parameters. The simulation study showed that we expect to be able to identify $p_3$ and $p_4$ using data that is available to ecosystem modelers via NEON. More specifically, in Figure \ref{fig:DC_plot} (top row) we show that the posterior narrows as $r$ increases for $p_3$ -- that is, it is identifiable with the observed flux data. We can also see that without additional flux data to help constrain $p_4$, the parameter was identifiable but non-estimable (INE) (Figure \ref{fig:DC_plot}, middle row). Although the uncertainty for both $p_3$ and $p_4$ is considerably lower in the NEON $GPP$ case (Figure \ref{fig:DC_plot}, right two columns), the parameters are not centered near the true values. The NEON $GPP$ case does show that $p_{11}$ can be estimated from data, but it is not estimable with only the NEON $NEE$ data.    

 As $r$, the number of times the data is cloned, was increased to 25 replicates, the posterior estimates for $p_4$ started to become more similar. However, it was not centered around the true values used for the simulation, which may be due to a lack of flux data available to constrain them. $p_4$ is likely to be INE - it is identifiable, but is non estimable given the collected data. $p_{11}$ continues to be tightly centered around the true value for the NEON $GPP$ case when as $r$ increased. 



 The results of our data cloning analysis demonstrate that NEON flux data will require additional flux observations in order to estimate both the $p_3$ and $p_4$ parameters. Posterior estimates of these parameters from the initial dataset are quite different, with the NEON $GPP$ case having comparatively tighter estimates. A potential cause for these differences may be due to NEON $GPP$'s ability to estimate $p_{11}$ well from the start and not getting stuck near a local minimum. As the number of data cloning replicates increases, the posteriors begin to look very similar. Under the NEON $GPP$ case, $p_{11}$ is identifiable, while with flux data from NEON $NEE$, $p_{11}$ appears to be identifiable but non estimable, with the posterior remaining relatively uniform even with $r = 25$ replicates. Our findings illustrate one of the shortfalls of using data cloning: though we are able to determine whether parameters are estimable or identifiable, we cannot be sure that our chains have converged to the true value.

 
 \vspace{-0.5cm}

\section{Discussion}

Estimating the posterior distributions of parameters in multi-state state space models can be challenging when observations of the states are not readily available.  This is especially apparent in ecological models where states are often sampled at more coarse intervals than the model time-step, resulting in many latent states without direct constraint from corresponding observations.  Here we introduce a method for changing the timestep used for generating latent states in  processes model so that it is coarser than the operational timestep of the process model (i.e., timestep of the difference equations).  Our analysis revealed a large increase in the efficiency of estimation of the posterior distributions of the process precision parameters when using a coarser time-step of latent state estimation, while producing latent state estimates that achieved the nominal coverage rate. One strength of the approach we present is that it preserves the operational time of the process model used to simulate the ecosystem dynamics.  As a result, no adjustments were required to the process model.  Another strength is that the equations used to change the timestep of the latent states do not require the new timesteps to be equally spaced, giving the flexibility to allow the latent states of the model to operate on any time scale.  However, changes to the latent state timestep do influence the interpretation of the process uncertainty parameters because they represent the distribution of process error that propagates from one latent state to the next - longer time-intervals between latent states will likely lead to process error distributions with larger variance.

Beyond data gaps in time, gaps in data where particular states and/or fluxes are never observed presents challenges in the ability to estimate the posterior distributions of parameters (identifiablility). To examine identifiablility in the focal ecosystem model (DALECev), we confirmed that we were able to successfully recover process parameters and process precisions when all states and fluxes were observed at all time steps and then in the case where there were annual temporal gaps in the observations of states.  This indicates that under ideal data collection the parameters are identifiable using the approach presented here. However, a lack of identifiability occurred when a subset of the flux data was not available to constrain model parameters, as is the case in applications using real observations.  In this case, our approach had difficulty recovering multiple process model parameters in the DALECev model. In particular, $p_2, p_3,$ and $p_4$ were difficult to estimate without all of the related fluxes used to constrain them.  These parameters govern the proportional allocation of photosynthesis (GPP) to respiration, foliage, and roots (Equations \ref{eq:Cft} -- \ref{eq:Crt}), thus requiring observations of their individual production in order to constrain the individual parameters.  
 
Our inference about identifiability of process parameters was based on the application of data cloning. Other methods of assessing identifiability were considered, including Hessian methods \citep{viallefont, Little2009} and symbolic algebra methods \citep{ColeHMM, ColeSSM}. Hessian methods and symbolic algebra methods were deemed unfit due to the time series of the model being long, which would lead to problems with numerical stability using the Hessian method \citep{Bulla2008ComputationalII} and lack of computational resources to perform the symbolic algebra calculations in MAPLE \citep{ColeHMM}. Identifiability is a problem that has long plagued ecological and biological modeling \citep{LUO2009}, and data cloning is a simple method that can be used with simulated data prior to the design of an experiment to assist the design data collection schemes that mitigate identifiability challenges. Our simulation study used data cloning with observed flux data that would be available from NEON, and showed that additional flux data is required to constrain a subset of model parameters. It also illustrated that while some parameters are shown to be identifiable through data cloning, they are not necessarily centered at the true parameter values. The types of data measured at a NEON site is not atypical for a terrestrial ecosystem study, particularly those in the Ameriflux and Fluxnet networks. With simulation results showing that some parameters are not identifiable or identifiable but non-estimable, it is crucial that scientists have access to methods to help them assess whether they can trust the results they obtain from their modeling framework.

Data cloning has other uses aside from assessing identifiability and aiding in experimental data collection for ecosystem modelers. A well documented problem with soliciting prior distributions for parameters in Bayesian analyses is that non-informative prior distributions on one scale may become highly informative prior distributions when transformed (see \cite{LELE2020} and references therein for a thorough treatment). These falsely non-informative priors can lead to biases in parameter estimation and prediction, in turn leading to incorrect decisions made by stakeholders and policy makers \citep{LELE2020}. Data cloning methods can be used to expose accidental biases introduced through using these priors that are non-informative on one scale, as the data cloning posterior will approach the maximum likelihood solution as $r$ increases, and maximum likelihood estimation is invariant to re-parameterizations.
  
Our study focused on the development and evaluation of methods, and sets the foundation for future work. First, while simulation studies with data synthesized from the DALEC model is necessary to test our methods, it is important to test the performance of these methods using real observations. 
Second, the results for latent state updates discussed here have been for the univariate case where covariance between states is not considered, though the states can be updated \textit{en bloc} with multivariate normal Gibbs updates. Multivariate latent state updates could be complemented with conjugate Gibbs updates for the covariance matrix, allowing full estimation of the covariance structure and (potentially) better latent state estimates. The methods presented here could be extended to non-linear Markovian process models by linearizing the update equations, though we have not presented any examples here. Third, the Gibbs updates shown in this paper are only applicable to state space models where both the observation and process model errors are normally distributed. While this is a common assumption in terrestrial carbon models, other applications may have error structures that do not meet this assumption. For example, error structures may be needed to maintain ecological realism, such as positivity of a particular latent state. More complex error structures or model dynamics require alternative fitting methods, such as particle methods \citep[see][for a thorough review]{KANTAS2014} or Gaussian Process Regression \citep{TURNER2010}. Finally, we have shown that it may not be possible to identify or recover process parameters for the DALECev model  under yearly data gaps when using only data available from NEON. However, it is likely that assimilating additional data not observed by NEON, such as satellite-derived leaf area index (e.g., MODIS LAI), and incorporating stronger priors that reflect general ecological principles will help to constrain model parameters further \citep{BLOOM}.
 
In conclusion, to address the growing popularity of state space modeling in the ecological forecasting research, we propose methods that help to assess and fix problems with process precision estimation and identifiability of process parameters that frequently arise in ecosystem state space modeling when observations are scarce. The state space framework augmented with data cloning to assess identifiability of parameters presented here is flexible enough to be adapted and applied to a broad range of problems including non normal-normal error structures, non-linear process models, and spatio-temporal models. The methods discussed here will allow practitioners to more effectively and efficiently address and overcome common suites of problems that arise when using state space models. 


\bibliography{dalec}
\bibliographystyle{apalike}
\end{document}